\documentstyle[11pt,newpasp,twoside,epsf,epsfig]{article}
\def\vev#1{\left\langle #1\right\rangle}
\def\edcomment#1{\iffalse\marginpar{\raggedright\sl#1\/}\else\relax\fi}
\reversemarginpar

\begin{document}
\title{The Origin and Distribution of Angular Momentum in Galaxies}
\author{Joel R. Primack}
\affil{Physics Department, University of California, Santa Cruz, CA
95064 USA}

\begin{abstract}
Cold Dark Matter with a large cosmological constant ($\Lambda$CDM)
appears to fit large scale structure observations well.  Of the
possible small scale problems, the {\it Central Cusps} and {\it Too
Many Satellites} problems now appear to be at least partly solved, so
{\it Angular Momentum} has become the most serious remaining CDM
problem.  There are actually at least two different angular momentum
problems: A.  Too much transfer of angular momentum to the dark halo
to make big disks, and B.  Wrong distribution of spedific angular
momentum to make spiral galaxies, if the baryonic material has the
same angular momentum distribution as the dark matter.  The angular
momentum of dark matter halos, and presumably that of the galaxies
they host, appears to arise largely from the orbital angular momentum
of the satellites that they accrete.  Since the dark and baryonic
matter behave very differently in such accretion events, it is
possible that the resulting angular momentum distribution of the
baryons is different from that of the dark matter, as required to make
the sort of galactic disks that are observed.  The latest
hydrodynamical simulations give some grounds for hope on this score,
but much higher resolution simulations are needed.
\end{abstract}

\section{Too Much Transfer of Angular Momentum to the Dark Matter to 
Make Big Disks}

\subsection{History}

This phenomenon was described by Navarro \& Benz (1991) and Navarro \&
Steinmetz 1997.  It appears to be due at least in part to the
unphysically rapid cooling of gas when feedback from star formation is
neglected.  When gas is prevented from cooling before $z=1$ more
realistic disks form (Weil, Eke, \& Efstathiou 1998; Eke, Efstathiou, \&
Wright 2000), as also can happen with cooling when feedback is treated
in simple ways (Thacker \& Couchman 2001, Maller \& Dekel 2002).  More
realistic disks may form even without feedback in higher resolution
simulations (Governato et al. 2002), although even with feedback many
of the baryons still formed a big bulge in one simulation (Abadi,
Navarro, Steinmetz, \& Eke 2003).  The merging history and the nature of
the assumed feedback evidently matter!

\subsection{Origin of the Angular Momentum Problem Transfer Problem:
Overcooling?}

Overcooling in merging satellites can explain why the baryons in
simulations lose much of their angular momentum to the dark matter.
If the baryons are more concentrated than the DM in the satellites,
the DM will be tidally stripped and the baryons will lose energy and
angular momentum by dynamical friction.  This is illustrated in Figure
1.  But if stellar feedback prevents the gas from cooling in small
accreted satellites, then the gas rather than the DM will be
preferentially stripped at large radius, and the gas may retain its
angular momentum.

\begin{figure}[htb]
\centering
\centerline{\epsfig{file=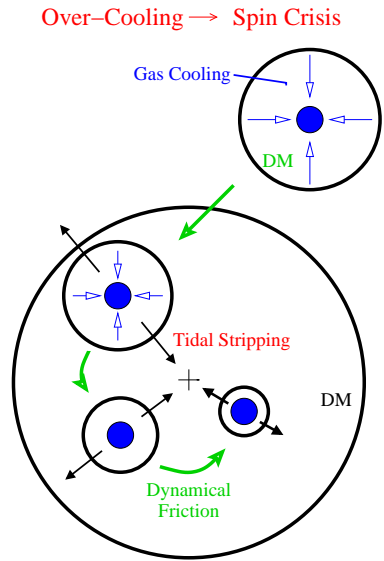,width=6cm}}
\caption{Overcooling model (Maller \& Dekel 2002).}
\end{figure}

\begin{figure}[h]
\plottwo{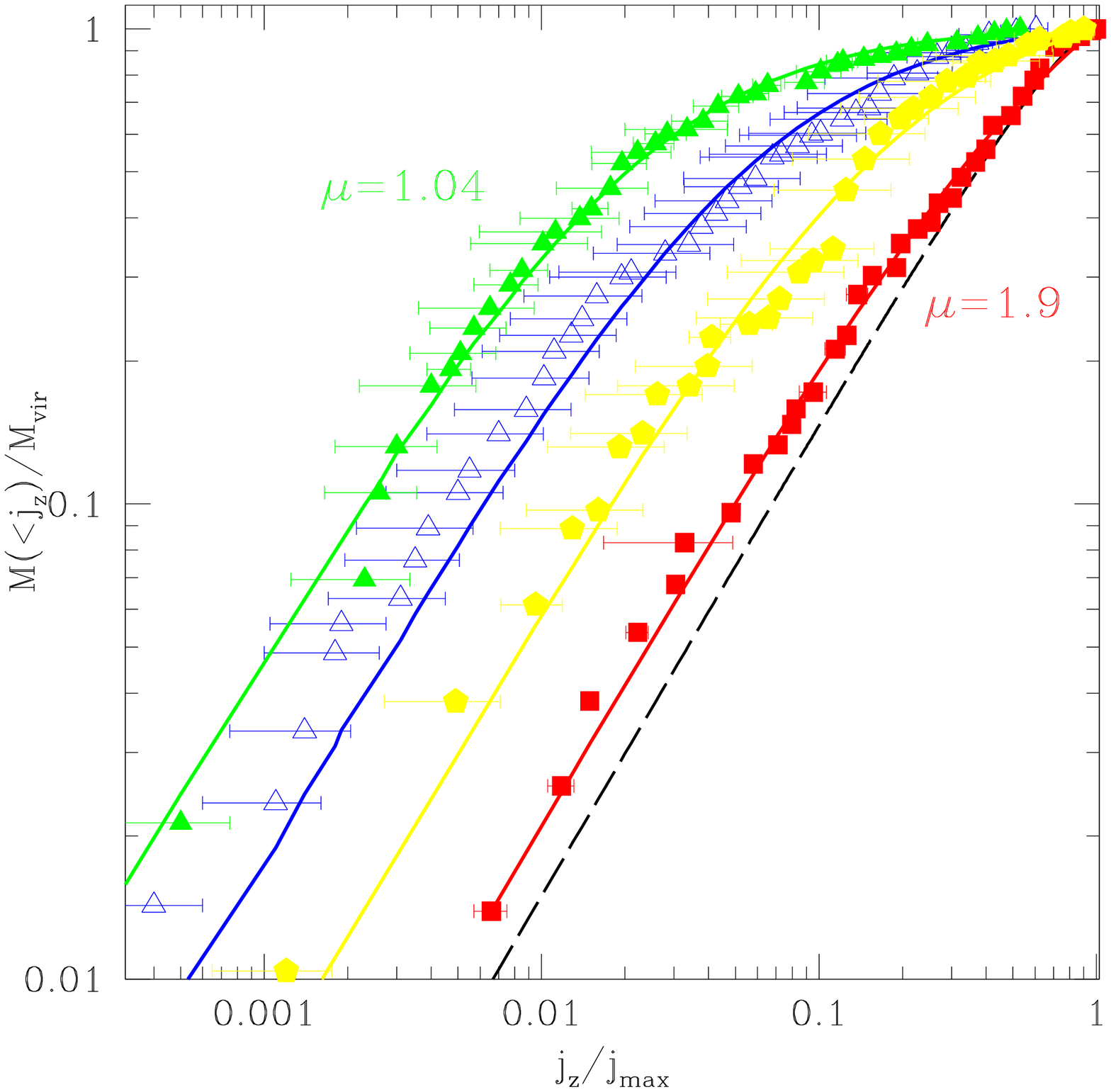}{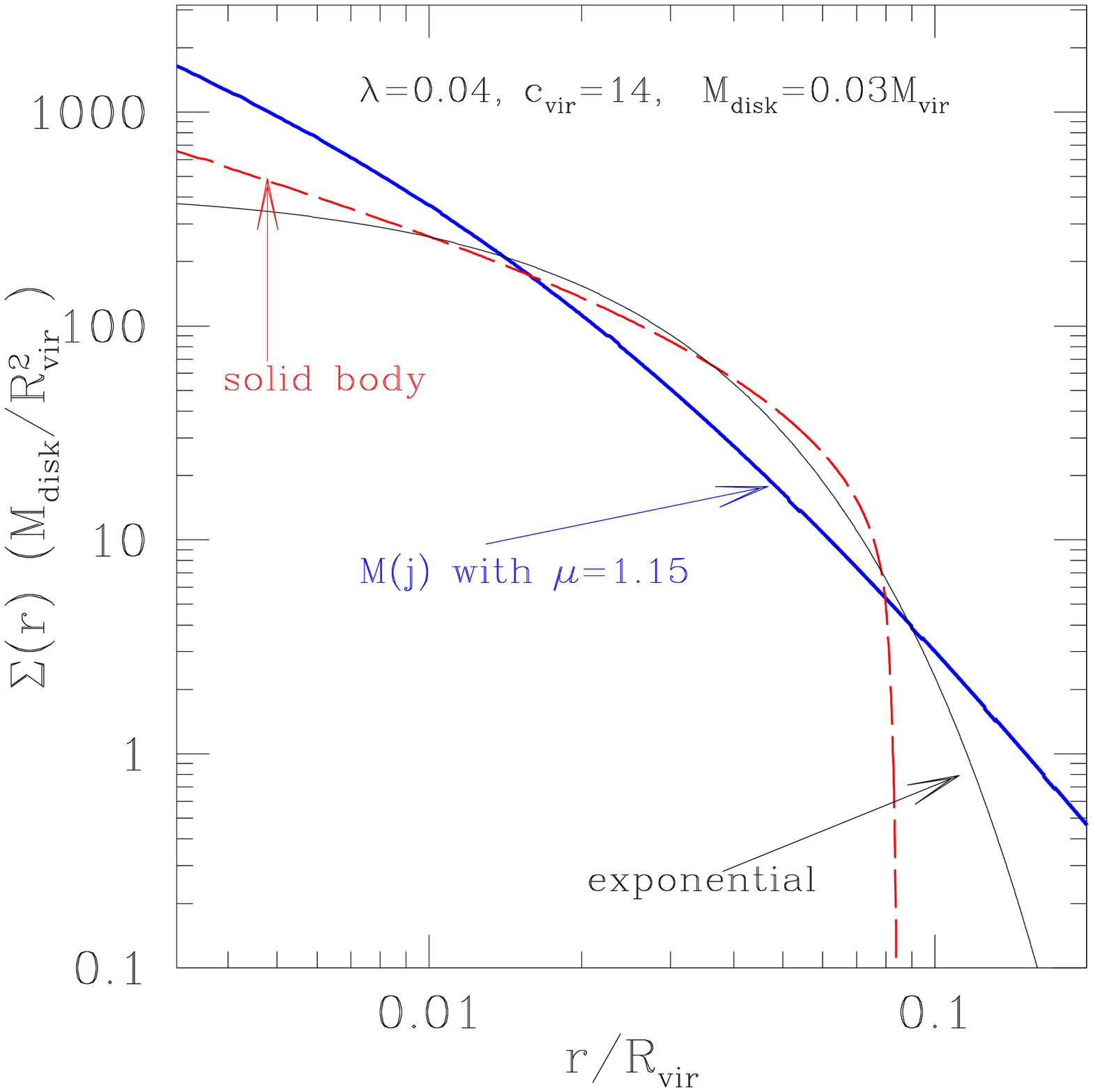}
\caption{(a) Mass distribution of angular momentum in four halos
spanning a range of $\mu$ values from 1.04 to 1.9.  The curves are the
funtional fits, eq. (1).  Profiles are normalized to coincide at
$M_v$, where $j=j_{\rm max}$.  The distribution for a uniform sphere
in solid body rotation is shown for comparision (dashed line). (b)
Typical disk surface-density profile implied by our universal $M(<j)$
distribution for typical $\mu=1.25$, assuming that $j$ is conserved
during baryonic infall (Mestel).  There is much more material at small
$r$ than in an exponential disk, and also a tail to large $r$. 
(Bullock et al. 2001.)}
\end{figure}

\section{Wrong Distribution of Specific Angular Momentum to Make 
Spiral Galaxies}

As part of James Bullock's dissertation research, we found that the
distribution of specific angular momentum $j$ in dark matter halos has
a universal profile, described by the equation below and illustrated
in Fig. 2(a), in which much of the DM has low $j$ but there is also a
high $j$ tail (Bullock et al. 2001):
\begin{equation}
M(<j) = M_v {{\mu j}\over{j_0 + j}} \ , \qquad \mu>1 \ .
\end{equation}
This profile has $j_{\rm max}=j_0/(\mu-1)$.  It is roughly power-law
for $j < j_0$ and flattens out for $j > j_0$ .  The quantity $\mu$ $(>
1)$ is a shape parameter: for $\mu \gg 1$, $M(<j)$ is a pure power
law, while for $\mu \longrightarrow 1$ only half the mass falls in the
power-law regime and there is a pronounced bend.  We find that
$\log(\mu-1)$ has a Gaussian distribution with mean 0.6 ($\vev{\mu} =
1.25$) and $\sigma = 0.4$. \footnote{Other groups (Chen \& Jing 2002;
Chen, Jing, \& Yoshikawa 2003) have confirmed our universal profile
$M(<j)$, although they find that some halos have a significant amount
of negative $j$ material (rotating in the opposite sense), which could
exacerbate the angular momentum profile problem.  This is probably
partly a resolution issue, and partly an issue of methodology. The
fraction of mass with $j$ negative is much larger using the Particle
method (as in van den Bosch et al. 2003ab) than using the Cell method
(as in Bullock et al. 2001).  But I do not understand the logic of the
Particle method applied to gas particles; if gas particles had the
large random momenta assigned in this approach, they would constantly
collide and shock.}

\subsection{Implications for Galactic Disks}

If the baryons that become the visible parts of galaxies have the same
angular momentum distribution as the DM, they could not form the
observed rotationally-supported disks.  This is illustrated in Figure
2(b).

It has long been assumed that baryons and DM in a halo start with a
similar angular momentum distribution, based on the idea that the
angular momentum arising from large scale tidal torques will be
similar across the entire halo.  But this may not be true if the
angular momentum of halos grows mainly from the orbital angular
momentum of accreted satellite halos, as we recently proposed
(Vitvitska et al. 2002).  We discuss this further below.

\subsection{Does Taking Bulges Into Account Help?}

Of course, some of the low $j$ baryons will form bulges. The mismatch
between the surface density profiles $\Sigma(r)$ is most severe for
halos with small $\mu \approx 1$, which also tend to have mis-aligned
angular momentum distributions and are therefore less likely hosts of
spiral galaxies.  But even for well-aligned halos with $\mu > 1.1$,
the implied central densities are still higher than in exponential
disks (Fig. 2b).  And assuming that all the low-$j$ material forms a
bulge (van den Bosch et al. 2003ab) results in $B/D$ ratios that are
too high, with almost no pure disk galaxies like M33.  Moreover, in
the previous figure we assumed that angular momentum is conserved.
Any transfer of angular momentum to the dark matter (as seen e.g. by
Navarro \& Steinmetz 1997) would exacerbate the problem.

\section{Solutions?}

Only for shape parameter $\mu \ga 2$, which occurs for fewer than
$10\%$ of halos, does the surface density $\Sigma(r)$ look like the
exponential of observed disk galaxies.  If CDM is right, then either
the baryons have a $j$ distribution rather different from that of the
dark matter, or only a special subset of the baryons in the halo forms
the disk, or both.

How do forming halos actually acquire their angular momenta?
Vitvitska et al. (2002) proposed and investigated the hypothesis that
it mostly comes from the orbital angular momentum of accreted
satellite halos.  If so, since the dark matter particles pass through
each other while the baryons shock during the hierarchical merging
process through which structure forms in CDM, then it would actually
be rather plausible that these two components will have a different
distribution of specific angular momentum $j$.  Such different
distributions could start to arise at high redshift (van den Bosch et
al. 2003ab).

\subsection{Origin of Angular Momentum $J$ in Dark Matter Halos}

The standard picture is that $J$ arises from tidal torques (Peebles
1969, Doroshkevish 1970, White 1984).  This is no doubt true at some
level, but this model incorrectly predicts both the amplitude (Barnes
\& Efstathiou 1987, Sugarman et al. 2000) and the direction (Lee \&
Pen 2000, Porciani et al. 2002).  An improved version of the tidal
torque model still predicts dependence of $\lambda'$ on mass,
redshift, and cosmology, in disagreement with simulations (Maller,
Dekel, \& Somerville 2002).

\subsection{Halo Mass Accretion Histories}

In her dissertation research with me, Risa Wechsler (Wechsler et
al. 2002) investigated the formation history of every large dark
matter halo in the same high-resolution cosmological simulation that
James Bullock had earlier analyzed.  The result was several thousand
structural merger trees, including the merging halos' structural
properties such as the radial distribution of the dark matter (NFW
concentration parameter) and the angular momentum (spin parameter).
This allowed us to explain how the concentration of a halo is related
to its mass accretion history (also subsequently studied by van den
Bosch 2002 and Zhao et al. 2003).  

\begin{figure}[t]
\centering
\centerline{\psfig{file=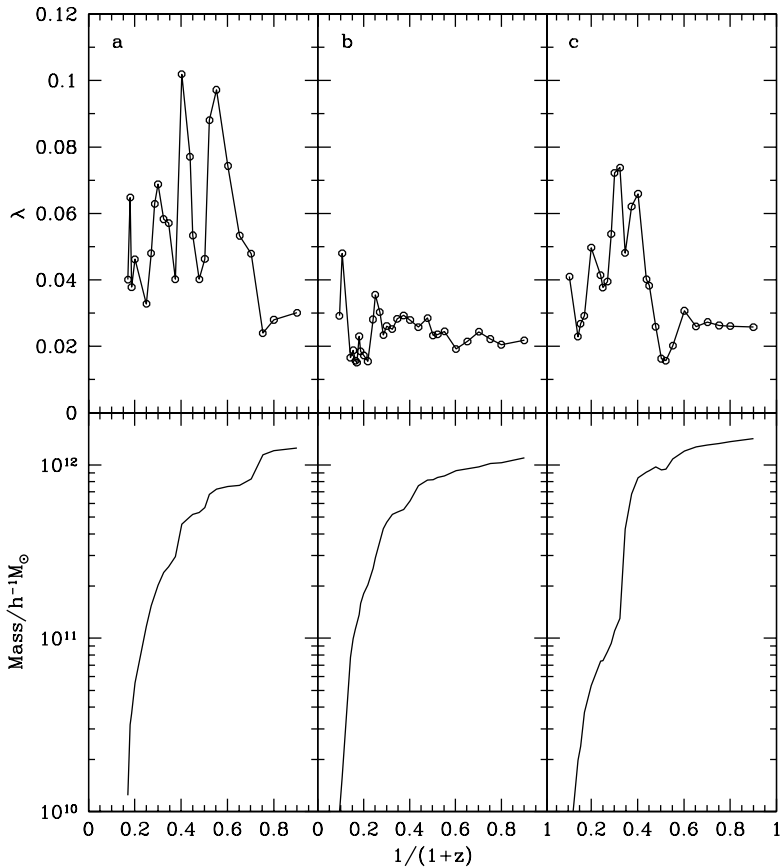,width=12cm}}
\caption{Mass accretion (lower panels) and spin parameter evolution
(upper panels) of three galaxy-mass halos.  Halos typically show fast
mass growth at high redshift with rapid changes of spin parameter,
followed by slower mass accretion with spin parameter usually
declining.  (From Vitvitska et al. 2002.)}
\end{figure}

Briefly, we found that 

\begin{itemize}

\item There is an early epoch of rapid mass accretion often involving major
mergers: the central region acquires a density similar to the background
density at that epoch.

\item A subsequent epoch of slow mass accretion by minor mergers mainly
builds the outer part of dark matter halo with $r > r_s$, increasing the
concentration of the halo: $C_v=R_v/r_s$.  

\end{itemize}
\noindent We are incorporating this information into semi-analytic
modelling of galaxy formation, where it has significant effects.

We also investigated how the angular momentum of halos is related to
their mass accretion histories; the results were reported in Vitvitska
et al. (2002).  We studied the evolution of three $10^6$ particle halos of mass
$\sim10^{12} M_\odot$, shown in Fig. 3 here, 
and also hundreds of halos in a large
simulation.  The spin parameters
\begin{equation}
\lambda \equiv {{J |E|^{1/2}}\over{G M^{5/2}} } \ , \quad 
\lambda' \equiv {j \over {\sqrt(2) M_{\rm vir} V_c R_{\rm vir} }} 
\end{equation}
typically have sharp increases (or sometimes decreases) due to major
mergers during the early rapid accretion phase of halo growth, and a
steady decline during the subsequent slower accretion.
Halos typically have rapid increases of $\lambda$ in major mergers if
$\lambda$ is small, but $\lambda$ decreases if $\lambda$ is large.
The spin parameter $\lambda$ typically shows big jumps due to major
mergers, and slow decline during the slow mass accretion epoch due to
random orientations of orbital angular momenta of accreted satellites.

\section{Random Walk Model of Angular Momentum Growth}

In order to model the angular momentum history of halos as due to
accretion in Vitvitska et al. (2002), we measured the velocity
anisotropy parameter of the incoming satellites, finding $\beta
\approx 0.6$, and also the dependence of of the satellite orbital
angular momenta on the satellite internal velocity.  We checked that
the incoming directions of satellites are essentially random, and used
the Extended Press-Schechter approximation to model halo mass
accretion histories.  We found that the resulting pattern of spin
parameter evolution is similar to that in high-resolution N-body
simulations.  We then calculated the $\lambda$ distribution in this
{\it random walk} model of halo angular momentum growth, and found it
to agree very well with N-body simulations.

Our random walk angular momentum accretion model agrees with N-body
simulations in the resulting spin parameter distribution (Vitvitska et
al. 2002; Maller, Dekel, \& Somerville 2002).  Maller \& Dekel (2002)
showed that a simplified version also leads to halos having the $j$
distribution seen in simulations.  It is interesting that the random
walk model predicts that halos that had a major merger have higher
$\lambda$.  This may lead to an elliptical galaxy paradox, and it
makes intersting predictions that can be compared to new data on the
rotation of halos hosting elliptical galaxies (see Vitvitska et
al. 2002).

\section{Possible Solutions to the Angular Momentum Distribution Problem}

\subsection{Solving Angular Momentum Problems Via Feedback}

It is the late mass accretion of small halos that generates much of
the low angular momentum material in big halos.  Maller \& Dekel
(2002) point out that in small halos supernova feedback is especially
effective in preventing gas from cooling, unlike in the Fig. 1
cartoon.  Consequently, the gas in small halos will be less
concentrated than the DM, so the gas is stripped before it reaches the
center of the halo.  They found that feedback characterized by a
parameter $V_{fb}=95$ km/s reduces gas content of dwarf galaxies to
observed levels, and also results in higher spin for dwarf galaxies
than giants, as observed (van den Bosch, Burkert, \& Swaters 2001).  
Massive galaxies are able to retain a much higher baryonic
fraction in this model, again in agreement with observations.  The
Maller \& Dekel (2002) feedback model with $V_{fb}=95$ km/s predicts
disk mass fractions for dwarf galaxies in good agreement with
observations, and well below the typical DM $j$ distribution at
both low and high $j$.

\subsection{Some Alternative Solutions to $j$ Distribution Problem}

Note that most of the visible baryonic mass is in spheroids with low
angular momentum, and that most of the baryonic mass does not end up
in the visible parts of galaxies.  What is needed to make observed
disks is for them to form from the baryons with the {\it right}
angular momenta.

This could happen if the low angular momentum material forms a central
concentration which is then blown out because of a starburst. But
is this likely in dwarfs, for which the angular momentum distribution
problem is most severe?

An alternative suggested by Katz et al. (2003) and by Birnboim \& Dekel
(2003) is that the baryonic material in most disk galaxies is fed in
rather cold (without virial shocks) from filaments, as mentioned by
Françoise Combes in her talk at this meeting. But it is unclear why
the angular momenta of the accreted gas should be coherent enough to
solve the angular momentum distribution problem.

The most optimistic alternative is perhaps that discussed by Barnes
(2002), which suggests that in hydrodynamic simulations of major
mergers of disk galaxies, the tidal tails have enough baryonic
material with large enough angular momentum to build realistic disks.
However, this was based on simulations in which the gas was assumed to
be isothermal; my group is finding that it may not be typically true in
more realistic simulations including shock-heating of the gas (Cox et
al. 2003).

\begin{figure}[ht]
\centering
\centerline{\epsfig{file=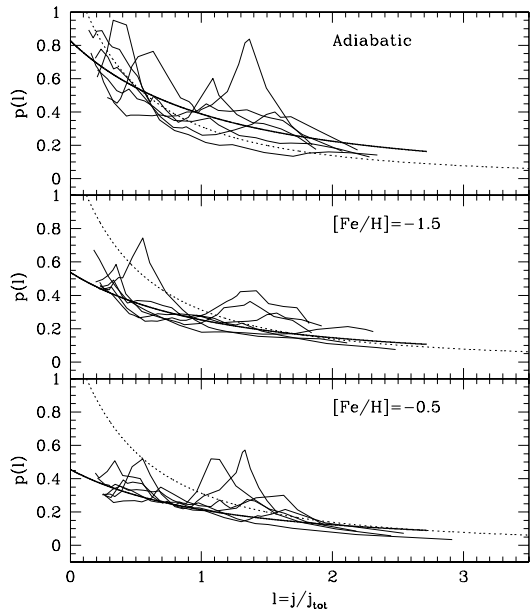,width=10cm}}
\caption{Angular momentum distribution of gas in nonradiative cases
(upper panel) and the hot gas fraction of all gas in cooling
simulations (lower two panels; the lower, higher metallicity case
cools faster).  Here $l=j/j_{\rm tot}$, where $j_{\rm tot} \equiv \vev{j}$
of hot gas.  The dotted and solid lines represent $\mu=1.25$ (typical dark
matter value) and $\mu=1.8$ (needed to form a roughly exponential
disk). (From Chen, Jing, \& Yoshikawa 2003.)}
\end{figure}

\section{Hydro Simulations: Do Baryons Have a Different $j$ Distribution?}

The first hydrodynamic simulations including cooling to address the
angular momentum distribution problem have just been reported by Chen,
Jing, \& Yoshikawa (2003).  They find that the hot gas has less
negative $j$, and a distribution of angular momentum that is much more
favorable for disk formation than that of the dark matter or the cold
gas.  Some examples are shown in Fig. 3.  However, the halos in their
simulation with adequate resolution are group-mass systems.  This
would be good news for disk galaxy formation if, in galaxy-mass halos,
much of this hot gas has enough time to cool so that it can make
galactic disks that are as old as they are typically observed to be --
which remains to be seen.

The shape parameter $\mu$ is much higher for hot gas than dark matter
in these hydro simulations.  (Recall that $\mu \ga 2$ is needed to get
an approximately exponential disk.)  {\it If} the hot gas in these
group-size halos corresponds to the gas that forms disks in typical
galaxies, this suggests that the gas will be able to form realistic
disks, even if it loses some of its angular momentum.

\section{Conclusions, Caveats, and Outlook}

\begin{description}
\item [$\bullet$] The Maller \& Dekel (2002) model may capture main
cause of the angular momentum transfer problem, and it suggests that
inclusion of feedback may solve it.

\item [But] this model is oversimplified and thus far only beginning
to be tested by hydro simulations (van den Bosch et al. 2003ab).

\item [$\bullet$] The random walk model of the origin of halo angular
momentum from the orbital angular momentum of accreted satellites
(Vitvitska et al. 2002) fits N-body simulation results.  It offers
hope that gas shocking may lead to the baryons having a different
specific angular momentum $j$ distribution from the dark matter, as
required to form galactic disks like those observed.

\item [But] thus far hydro simulations have not found this for all the
gas, and {\it most importantly} for the gas that actually forms
galactic disks.

\item [$\bullet$] Chen, Jing, \& Yoshikawa (2003) find that the hot
gas in group-size halos does have the sort of $j$ distribution that is
needed to form exponential disks.

\item [But] it is not clear that this hot gas has anything to do with
galactic disks.  Higher resolution hydrodynamic simulations are
clearly needed.

\item [$\bullet$] Since only part of the gas in halos forms the visible
material in galaxies, we can solve the $j$ distribution problem by
choosing the {\it right} gas.

\item[But] so far we have little understanding of whether and why
nature may do so.  This is perhaps the biggest challenge for CDM!

\end{description} 

\noindent {\it Acknowledegments:} My understanding of the topics
discussed here owes a great deal to discussions with James Bullock,
Avishai Dekel, Sandra Faber, Neal Katz, Anatoly Klypin, Andrey
Kravtsov, Ari Maller, Rachel Somerville, and Risa Wechsler.  Our
research is supported by grants from NASA and NSF at UCSC and NMSU,
and we have used simulations performed at the DOE National Energy
Research Scientific Computing Center (NERSC) headquartered at LBNL.
Finally, I must thank Ken Freeman and Mark Walker for organizing such
an interesting symposium and inviting me to give this talk.

\end{document}